\documentclass[12pt]{article}
\usepackage{amsmath,amssymb,bm,graphicx}
\usepackage{color} 

\begin{document}

\begin{flushright}
\end{flushright}


\begin{center}
{\Large{\bf  A new magnetic monopole inspired by Berry's phase 
}}
\end{center}
\vskip .5 truecm
\centerline{\bf Shinichi Deguchi~$^1$
 {\rm and} Kazuo Fujikawa~$^2$ }
\vskip .4 truecm
\centerline {\it $^1$~Institute of Quantum Science, College of 
Science and Technology}
\centerline {\it Nihon University, Chiyoda-ku, Tokyo 101-8308, 
Japan}
\vskip 0.4 truecm
\centerline {\it $^2$~Interdisciplinary Theoretical and Mathematical Sciences Program, 
}
\centerline {\it   RIKEN, Wako 351-0198, 
Japan}

\vskip 0.5 truecm

\makeatletter
\makeatother

\begin{abstract}
A new  static and azimuthally symmetric  magnetic monopolelike object, which looks like a Dirac monopole when seen from far away but smoothly changes to a dipole near the monopole position and vanishes at the origin, is discussed. This monopolelike object  is inspired by an analysis of an exactly solvable model of Berry's phase in the parameter space.    A salient feature of the monopolelike potential ${\cal A}_{k}(r,\theta)$ with a magnetic charge $e_{M}$ is that the Dirac string is naturally described by the potential ${\cal A}_{k}(r,\theta)$,  and  the origin of the Dirac string and the geometrical center of the monopole  are displaced in the coordinate space.
The smooth topology change from a monopole to a dipole takes place if the Dirac string, when coupled to the electron, becomes unobservable by satisfying the Dirac  quantization condition. The electric charge is then quantized even if the monopole changes to a dipole near the origin. In the transitional region from a monopole to a dipole, a half-monopole with a magnetic charge $e_{M}/2$ appears. 

\end{abstract}

 
\section{Introduction}
The magnetic monopole has a long history starting with the proposal of Dirac~\cite{Dirac} and among the past works, the mathematical clarification by Wu and Yang  \cite{Wu-Yang} is important for an analysis of a new monopolelike object in the present paper.  We have recently encountered an interesting monopolelike object \cite{Deguchi-Fujikawa1} in the analysis of an exactly solvable model \cite{exact solution}  of Berry's phase \cite{Higgins, Berry, Simon}. The monopolelike object  associated with Berry's phase is defined in the parameter space described by the external magnetic field but here we shall present a  translation of the monopolelike object  to the one in the real space. This monopolelike object shows several  novel properties. A notable feature of the new magnetic monopolelike potential ${\cal A}_{k}(r,\theta)$  with a magnetic charge $e_{M}$ is
that the Dirac string, which  appears when the net outgoing flux is nonvanishing, is naturally described by the potential ${\cal A}_{k}(r,\theta)$ itself and that the geometrical center of the monopole configuration and the origin of the Dirac string are displaced in the coordinate space.   We discuss the smooth topology change from a monopolelike configuration to a dipole configuration by combining this displacement with Gauss' theorem
$\int_{S}(\nabla\times {\cal A})\cdot d\vec{S}=\int_{V} \nabla\cdot (\nabla\times {\cal A}) dV=0$ for a volume $V$ which is free of singularities. Gauss' theorem shows that the monopolelike configuration  combined with the Dirac string is essentially a dipole. The smooth topology change from a monopole to a dipole then takes place when one regards the Dirac string as unobservable if the Dirac string  satisfies the Wu-Yang gauge invariance condition or equivalently Dirac's quantization condition when coupled to the electron, and thus it is related to the very idea of Dirac string. The quantization of the electric charge takes place even if the monopole changes to a dipole near the origin. In the transitional region from a monopole to a dipole, a half-monopole with a magnetic charge $e_{M}/2$ appears depending on the electric charge quantization condition.  
To our knowledge, no previous analysis of this class of deformation of the Dirac monopole in the real space has been performed. 

\section{A new static monopolelike potential}

\subsection{Basic properties of the conventional monopole}
We first briefly recall the characteristic properties of the conventional monopole, namely, topology and the quantization of the electric charge.  The standard static spherically symmetric monopole located at the origin is defined by the potential \cite{Wu-Yang}
\begin{eqnarray}\label{conventional potential}
{\cal A}_{\varphi}(r,\theta) 
= \frac{e_{M}}{4\pi r\sin\theta} \left(1 - \cos \theta \right)
\end{eqnarray}
and ${\cal A}_{\theta} ={\cal A}_{r} =0$ with a magnetic charge $e_{M}$.  This potential has a singularity at $\theta=\pi$ which is related to the Dirac string in addition to the singularity at $r=0$. 
This potential for $\theta \neq \pi$ and $r\neq 0$  gives rise to the magnetic field
\begin{eqnarray}\label{naive monopole}
\vec{\nabla}\times \vec{{\cal A}}=\frac{e_{M}}{4\pi r^{2}}\frac{\vec{r}}{r}
\end{eqnarray}
where the singularity explicitly appears only at $r=0$. We then have naively
\begin{eqnarray}\label{magnetic flux0}
\int_{S}(\nabla\times {\cal A})\cdot d\vec{S}=e_{M}
\end{eqnarray}  
by closing the excluded point $\theta=\pi$ on a sphere $S$ which covers the monopole located  at the origin. 
This value is independent of the size of the radius of a sphere which covers the monopole, namely, irrespective of a large radius far away or  a very small radius around the origin.  This is one of the topological aspects of the conventional monopole.

To satisfy the constraint of Gauss' theorem
$\int_{S}(\nabla\times {\cal A})\cdot d\vec{S}=\int_{V} \nabla\cdot (\nabla\times {\cal A}) dV=0$ for a volume $V$, that is free of singularities at $\theta=\pi$ and $r=0$,  Dirac introduced an infinitely thin tube surrounding the Dirac string  and stretching from the origin to infinity in the direction of $\theta=\pi$. This string or tube carries in the magnetic flux to the center of the magnetic monopole to compensate for the outgoing radial flux. If one chooses the volume $V$ such that it avoids this thin tube, i.e., the Dirac string,  inside a sphere with radius $r$ and if one chooses $S$ to be the entire surface of $V$, one can satisfy Gauss' theorem and the conservation of the magnetic flux. 

Dirac then showed that this thin tube can be excluded from the physical state space if one imposes a condition which leads to the electric charge quantization      
\begin{eqnarray}\label{Dirac quantization condition0}
\frac{e e_{M}}{c\hbar}=n\times 2\pi
\end{eqnarray}
with an integer $n$.  Here $e$ stands for the charge of the electron. After this elimination of the Dirac string from physical states, only the naive magnetic monopole with the flux in \eqref{naive monopole} and  \eqref{magnetic flux0} is left as the physical entity. This quantization is a remarkable prediction arising from the analysis of the Dirac string associated with the magnetic monopole \cite{Dirac}.  Wu and Yang explained the quantization of the electric charge using a different mathematical formulation \cite{Wu-Yang}.

    The monopolelike configuration we are going to discuss, which is inspired by an analysis of Berry's phase \cite{Deguchi-Fujikawa1}, is schematically shown for the case $r/a >1$ with a constant parameter $a\neq 0$ which sets the scale in Fig.1. The property of the monopolelike potential is sensitive to $r$, namely, at which value of $r$ the monopolelike object is observed. For example, for $r/a \rightarrow \infty$ the monopolelike object looks like the conventional monopole, while for $r/a \rightarrow 0$ with fixed $a\neq 0$, no singularity appears near the origin $r \sim 0$ and a dipole is realized; in fact it vanishes at $r=0$.  In this analysis, the coordinate origin, which has no meaning as physical reality, is identified as the location of the monopole on the basis of  the observation of the magnetic flux at  far away points. 
In contrast, the charge quantization takes place  in our monopolelike potential also, since the monopolelike potential agrees with the conventional monopole with a Dirac string when it is observed at any far away point and thus the conventional Dirac quantization is required to be satisfied to  ensure the unobservable Dirac string at such far away points. The conventional monopole \eqref{conventional potential} is recovered by setting $a = 0$ first with any fixed $r$.

\begin{figure}[htb]
 \begin{center}
   \includegraphics*[width = 6.0cm]{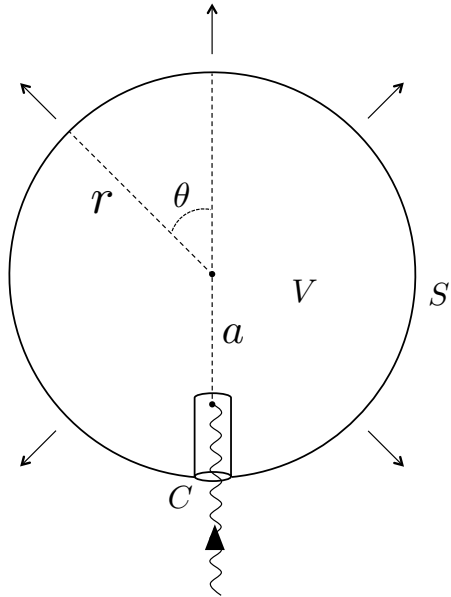}
   \caption{Schematic picture of a monopolelike object we discuss, covered by  a virtual sphere  with fixed radius $r$ which satisfies $r/a >1$ for a constant  parameter $a\neq 0$.   Geometrical center and the origin of the Dirac string  denoted by a wavy line are displaced by the distance $r=a$. The outgoing arrows indicate the outgoing flux in the radial direction on the sphere with fixed $r$.  Note that the dot at the geometrical center is  for a visual aid and does not represent any singularity at the origin $r = 0$ for fixed $a \neq 0$; in fact, the monopolelike object vanishes at $r = 0$. Volume $V$ inside the sphere, which is used later, avoids a thin tube surrounding the Dirac string and $S$ stands for the entire surface of $V$ including the surface of the tube in this context.} \label{Fig_01}
 \end{center}
\end{figure}

\subsection{New monopole-like potential} 
   
We now discuss a new static and azimuthally symmetric magnetic monopolelike potential defined by 
\begin{eqnarray}\label{potential1}
{\cal A}_{\varphi}(r,\theta) 
\equiv \frac{e_{M}}{4\pi r\sin\theta} \left(1 - \cos \Theta(\theta,\eta) \right)
\end{eqnarray}
and ${\cal A}_{\theta} ={\cal A}_{r} =0$ with a magnetic charge $e_{M}$.  This is a  translation of a monopolelike object associated with Berry's phase in the parameter space \cite{Deguchi-Fujikawa1} to the one in the real space. The variable $\Theta(\theta,\eta)$ is defined by
\begin{eqnarray}\label{capital theta}
\Theta(\theta,\eta)=\theta-\alpha(\theta,\eta)
\end{eqnarray}
where $\alpha(\theta,\eta)$ is defined by the relation
\begin{eqnarray}\label{parameter}
\tan\alpha(\theta,\eta)=\frac{\sin\theta}{\eta + \cos\theta}
\end{eqnarray}
with 
\begin{eqnarray}\label{parameter2}
\eta=\frac{r}{a}.
\end{eqnarray}
Note that $\alpha(\theta,\eta)=0$, namely, $a=0$ for any fixed $r$ gives the conventional Dirac monopole. 
The definition of $\eta= r/a$ is a new ingredient of the present model in the real space. 
The constant $a$ is a free parameter with the dimensions of length which sets the scale of the monopolelike object. This constant $a$ replaces the period parameter in Berry's phase \cite{Deguchi-Fujikawa1} and
leads to a very different physical interpretation of the monopolelike potential. To be precise, $a$ is treated as a constant parameter in the present model, while $a$ being the inverse of the period is a dynamical variable in Berry's phase. Thus the important notions of adiabatic and nonadiabatic movements in Berry's phase do not appear in the present model.  
The magnetic charge $e_{M}$, which is quantized to be $2\pi\hbar$ in the case of Berry's phase \cite{Deguchi-Fujikawa1}, is chosen as a free parameter at this moment and later quantized in combination with the electric charge by the Dirac quantization condition.  Physically, the infinitely strong external magnetic field  in Berry's phase corresponds to the long distance behavior of the monopole in the real apace and the vanishingly weak external magnetic field in Berry's phase corresponds to the short distance behavior of the monopole in the real apace, and thus very different.

The conditions \eqref{capital theta} and \eqref{parameter}, which may look rather ad hoc in the context of a monopole in the real space, automatically appear in the context of Berry's phase and describe the deformation of the monopole leading to the smooth topology change from a monopole to a dipole \cite{Deguchi-Fujikawa1}. We find the mathematical properties of the monopolelike object defined in the real space illuminating. We thus present the mathematical analyses of the model defined by \eqref{potential1}, \eqref{capital theta}, \eqref{parameter} and \eqref{parameter2}
 below in the hope that they will interest the wider audience who are familiar with the conventional Dirac monopole. Naturally, the mathematical descriptions (and quoted figures) are closely related to the analyses of  the exact solution of Berry's phase \cite{Deguchi-Fujikawa1}.
 But the analysis of the  monopolelike object in the real space clarifies some important aspects such as the natural appearance of the Dirac string in the context of the monopole in the real space and the physical relevance of the Dirac string in the context of  the monopolelike object   associated with Berry's phase.   

We start with the analysis of the parameter $\alpha(\theta,\eta)$.
In the relation \eqref{parameter} between $\theta$ and $\tan\alpha(\theta,\eta)$ for the case $0\leq \eta<1$, 
we have a singularity at $\cos\theta_{0}=-\eta$ in the denominator of \eqref{parameter}. But this does not give rise to a singular relation between $\alpha(\theta,\eta)$ and $\theta$; one can confirm
\begin{eqnarray}\label{slope}
\frac{\partial\alpha(\theta,\eta)}{\partial\theta}=\frac{1+\eta\cos\theta}{(\eta+\cos\theta)^{2}+\sin^{2}\theta}
\end{eqnarray}
and thus 
\begin{eqnarray}\label{smoothness}
\frac{\partial\alpha(\theta,\eta)}{\partial\theta}|_{\theta=\theta_{0}}=1
\end{eqnarray} 
for $\cos\theta_{0}=-\eta$.
For the parameter range $\eta\geq 1$,  the relation \eqref{parameter} is smooth and for $\eta=1$, we have an exact relation
\begin{eqnarray}\label{half value}
\alpha(\theta,\eta=1)=\theta/2.
\end{eqnarray}
 For other parameter values, we  have  \begin{eqnarray}\label{limiting form}
&&\alpha(\theta,\eta)=\frac{1}{\eta}\sin\theta  \ \ \ \ {\rm for} \ \eta \gg 1,\nonumber\\
&&\alpha(\theta,\eta)=\theta -\eta \sin\theta  \ \ \ \ {\rm for} \ 0\leq \eta \ll 1.
\end{eqnarray}

It is later shown that the parameter domain $\eta=r/a>1$  implies the existence of a monopole configuration, to be precise, both the Dirac string and the net outgoing flux appear.  The domain $0\leq \eta=r/a<1$  implies the appearance of a dipole configuration (and the disappearance of a monopole configuration), namely, neither Dirac string nor net outgoing flux appear. The transition from $\eta>1$ to
$\eta<1$ through the critical value $\eta=r/a=1$ is thus important. 
For the parameters
$\eta=1\pm\epsilon$
with a small positive $\epsilon$, it is confirmed that the value $\alpha(\theta,\eta)$ departs from the common value $\frac{1}{2}\theta$ assumed at around $\theta=0$ and splits into two branches for the values of the parameter $\theta$ close to  $\theta=\pi$. We have $\alpha(\pi,\eta)=0$ for $\eta=1+\epsilon$ and  $\alpha(\pi,\eta)=\pi$ for $\eta=1-\epsilon$, respectively,  with the slopes
\begin{eqnarray}\label{singularity}
\frac{\partial\alpha(\theta,\eta)}{\partial\theta}|_{\theta=\pi} =\mp \frac{1}{\epsilon}
\end{eqnarray}
for $\eta=1\pm \epsilon$, respectively, using \eqref{slope}. We thus observe the singular jump for a small value of $\epsilon$ characteristic to the topology change in terms of $\alpha(\theta,\eta)$ at $\eta=1$. 

The parameter  $\Theta(\theta, \eta)=\theta-\alpha(\theta,\eta)$, which determines the monopolelike potential \eqref{potential1}, is shown in Fig.2. Note that $\Theta(0,\eta)=0$  since 
$\alpha(0,\eta)=0$, and 
\begin{eqnarray}\label{end values}
\Theta(\pi,\eta)=\pi, \ \pi/2, \ 0
\end{eqnarray}
respectively,  for $\eta>1$, $\eta=1$, and $\eta<1$, since 
$\alpha(\pi,\eta)=0, \pi/2, \pi$, respectively,  for these values of $\eta$. See, for example,  \eqref{half value} and \eqref{limiting form}.
We also have 
\begin{eqnarray}\label{turning point1}
\frac{\partial\Theta(\theta,\eta)}{\partial\theta}|_{\theta=\theta_{0}}=0
\end{eqnarray}
for $\eta<1$ using \eqref{smoothness}.

We here emphasize an important fact about the singularity structure at $r=0$ in the monopolelike potential
\eqref{potential1} for fixed $a\neq 0$.  From \eqref{limiting form} we have
\begin{eqnarray}\label{limiting form2}
\Theta(\theta,\eta)&=&\theta - \alpha(\theta,\eta)\nonumber\\
&\simeq&\eta \sin\theta 
\end{eqnarray}
for $0\leq \eta = r/a \ll 1$, namely, for $r\sim 0$ with fixed $a$. This shows that the monopolelike potential \eqref{potential1} behaves like
\begin{eqnarray}\label{singularity free}
{\cal A}_{\varphi}&\simeq& \frac{e_{M}}{8\pi r\sin\theta}(\frac{r}{a}\sin\theta)^{2} = \frac{e_{M}}{8\pi a^{2}}r\sin\theta
\end{eqnarray}  
at $r\sim 0$ with fixed $a \neq 0$. This behavior  shows that the monopolelike potential becomes {\em regular} with respect to both $r$ and $\theta$ near the origin and vanishes at the origin. This fact is important to understand the smooth topology change from a monopole to a dipole near the origin.  Also the dot at the geometrical center in Fig.1 is just for a visual aid and not a singularity,

\begin{figure}[htb]
 \begin{center}
   \includegraphics*[width = 8.0cm]{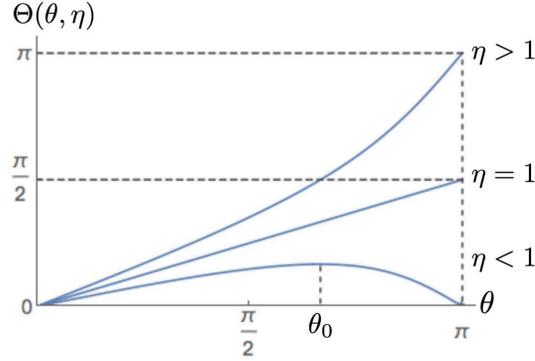}
   \caption{The relation between $\theta$
and $\Theta(\theta,\eta)$ parameterized by $\eta$. Note that $\cos\theta_{0}=-\eta$. The monopolelike potential \eqref{potential1} is determined by $\Theta(\theta,\eta)$. See Fig.3 in \cite{Deguchi-Fujikawa1}.}\label{Fig_02}
 \end{center}
\end{figure}

We have thus specified the monopolelike potential \eqref{potential1}.
The topology change is seen in the change of $\Theta(\pi,\eta)=\pi$ for $\eta>1$ to $\Theta(\pi,\eta)=0$ for $\eta<1$ in Fig.2.  But we have a well-defined exact potential at the boundary $\eta=1$
\begin{eqnarray}\label{half sphere}
{\cal A}_{\varphi}&=&\frac{e_{M}}{4\pi r\sin\theta}(1-\cos\frac{1}{2}\theta)
\end{eqnarray}  
for $\theta\neq \pi$. We also note that the Dirac string which corresponds to the singularity of the potential \eqref{potential1} can appear at $\theta=0$ or $\theta=\pi$; but no singularity at $\theta=0$ since $\Theta(0,\eta)=0$, and the possible Dirac string appears at $\theta=\pi$ for $\Theta(\pi,\eta)=\pi$ (for $\eta>1$) or $\Theta(\pi,\eta)=\pi/2$ (for $\eta=1$), but no string for $\Theta(\pi,\eta)=0$ (for $\eta<1$).

Using the potential \eqref{potential1} and ${\cal A}_{\theta}={\cal A}_{r}=0$,
we have  the magnetic flux in the space $\vec{r}=r(\sin\theta\cos\varphi, 
\sin\theta\sin\varphi,\cos\theta )$ for $\theta\neq \pi$, 
\begin{eqnarray}\label{magnetic flux}
\nabla\times {\cal A}
&=&\frac{e_{M}}{4\pi}\frac{\Theta^{\prime}(\theta,\eta)\sin\Theta(\theta,\eta)}{\sin\theta}\frac{1}{r^{2}}{\bf e}_{r} - \frac{e_{M}}{4\pi}\frac{\frac{\partial\Theta(\theta,\eta)}{\partial r}\sin\Theta(\theta,\eta)}{r\sin\theta}{\bf e}_{\theta}
\end{eqnarray}
where $\Theta^{\prime}(\theta,\eta)=\frac{\partial\Theta(\theta,\eta)}{\partial\theta}=1-\frac{\partial\alpha(\theta,\eta)}{\partial\theta}$ which is explicitly given using \eqref{slope}, ${\bf e}_{r}=\vec{r}/r$ and ${\bf e}_{\theta}$ is a unit vector in the (increasing) direction of $\theta$ in the spherical coordinates.
By recalling $\eta=r/a$, we have
\begin{eqnarray}
\frac{\partial\Theta(\theta,\eta)}{\partial r}=\frac{1}{a}\frac{\partial\Theta(\theta,\eta)}{\partial \eta}
=\frac{1}{a}\frac{\sin\theta}{(\eta+\cos\theta)^{2}+\sin^{2}\theta}
\end{eqnarray}
where we used $\Theta(\theta,\eta)=\theta-\alpha(\theta,\eta)$ and \eqref{parameter}.
Note that we have the standard monopole \eqref{naive monopole} in \eqref{magnetic flux}
\begin{eqnarray}\label{standard monopole}
\nabla\times {\cal A}
&=&\frac{e_{M}}{4\pi}\frac{1}{r^{2}}{\bf e}_{r}
\end{eqnarray}
for $\Theta(\theta,\eta)=\theta$ which is realized  for $\eta=r/a\rightarrow \infty$ or $a\rightarrow 0$.

As for the integrated net outgoing  flux in the radial direction from a sphere centered at $\vec{r}=0$, 
avoiding the singular point $\theta=\pi$, we have
\begin{eqnarray}\label{total flux}
\int_{\theta\neq\pi}(\nabla\times {\cal A})\cdot d\vec{S}&=&\int \frac{e_{M}}{4\pi}\frac{\Theta^{\prime}(\theta,\eta)\sin\Theta(\theta,\eta)}{\sin\theta}\frac{1}{r^{2}}r^{2}\sin\theta d\varphi d\theta\nonumber\\
&=&\int^{\pi}_{0} \frac{e_{M}}{2}\Theta^{\prime}(\theta,\eta)\sin\Theta(\theta,\eta) d\theta \nonumber\\
&=&\frac{e_{M}}{2} (1-\cos\Theta(\pi,\eta))
\end{eqnarray}
which agrees with Stokes' theorem applied to \eqref{potential1} for a small circle C near the south pole ($\theta=\pi$) in Fig.1.
Note that $\theta=\pi$ is a singularity of the potential but the point with $\theta=\pi$ on the surface is measure zero from a point of view of the present surface integral for the flux in the radial direction. 
The outgoing flux \eqref{total flux} specifies the typical topological configurations. In Fig.2, we have $\Theta(\pi,\eta)=\pi$  in the domain $\eta=r/a>1$, $\Theta(\pi,\eta)=\frac{1}{2} \pi$ in the transitional domain $\eta=r/a=1$, and $\Theta(\pi,\eta)=0$ in the domain $\eta=r/a<1$,
respectively. We thus have the integrated flux of a monopole
\begin{eqnarray}\label{integrated flux}
\int_{\theta\neq\pi} (\nabla\times {\cal A})
\cdot d\vec{S}&=&
e_{M},\ \ \frac{1}{2}e_{M},\ \ 0 
\end{eqnarray}
respectively, for three different parameter domains $\eta >1$, $\eta=1$ and $\eta<1$.

Our monopolelike potential is regular  for $\theta\neq\pi$ since both singular behavior \eqref{singularity} and the Dirac string  appear at $\theta=\pi$, and it is also regular at the point $r=0$ for $a\neq 0$ as explained in \eqref{singularity free}.

\section{Topological classification by Gauss' theorem}
 
The useful information about the topology change from a monopole for $r>a$ to a dipole for $r<a$ is given by Gauss' theorem which states that 
\begin{eqnarray}\label{Gauss}
\int_{S}(\nabla\times {\cal A})\cdot d\vec{S}=\int_{V} \nabla\cdot (\nabla\times {\cal A}) dV=0
\end{eqnarray}
using the formula of vector analysis $\nabla\cdot (\nabla\times {\cal A}) =0$ valid for the domain in which ${\cal A}$ is regular.
Here the volume $V$ is defined by excluding a thin tube covering the Dirac string, which is shown in Fig.1, and $S$ stands for the entire surface of this volume $V$.
Note that there is no singularity inside the volume $V$ by recalling that the origin $r=0$ is not a singular point for $a\neq 0$ \eqref{singularity free}.
The Dirac string originates at $z=-a$
on the negative z-axis
corresponding to $\eta=r/a=1$. Recall that no Dirac string appears for $\eta<1$
since $\Theta(0,\eta)=\Theta(\pi,\eta)=0$ as in Fig.2 and thus no singularity appears in \eqref{potential1} at either $\theta=0$ or $\theta=\pi$.

The discrepancy of \eqref{total flux} and \eqref{Gauss} is attributed to the contribution 
of the Dirac string.  
It is instructive to confirm Gauss' theorem \eqref{Gauss} for $\eta>1$.  The first term in \eqref{magnetic flux}
determines the contribution from the outer surface in Fig.1 
\begin{eqnarray}\label{outer surface}
\int_{S_{out}}(\nabla\times {\cal A})\cdot d\vec{S}=e_{M}
\end{eqnarray}
using the result in \eqref{integrated flux}.
The second term in \eqref{magnetic flux} describes a contribution to the flux flowing out from the volume V through the surface of the cylinder part of the thin tube surrounding the Dirac string in Fig.1
\begin{eqnarray}\label{cylinder}
\int (\nabla\times {\cal A})\cdot dS_{\theta}
&=& - \int \frac{e_{M}}{4\pi}\frac{\frac{\partial\Theta(\theta,\eta)}{\partial r}\sin\Theta(\theta,\eta)}{r\sin\theta}dr r\sin\theta d\varphi\nonumber\\
&=&-\frac{e_{M}}{2}\int_{a}^{r}
\frac{\partial\Theta(\theta,\eta)}{\partial r}\sin\Theta(\theta,\eta)dr\nonumber\\
&=&\frac{e_{M}}{2}(\cos\Theta(\theta,\eta)-\cos\Theta(\theta,\eta=1))\nonumber\\
&=&\frac{e_{M}}{2}(\cos\Theta(\theta,\eta)-\cos\frac{1}{2}\theta)
\end{eqnarray}
using  the surface element in the ${\bf e}_{\theta}$ direction
\begin{eqnarray}
dS_{\theta}=dr r\sin\theta d\varphi {\bf e}_{\theta}
\end{eqnarray}
and $\Theta(\theta,\eta=1)=\frac{1}{2}\theta$; we later set $\theta=\pi$.

As for the contribution of a small cap around the origin of the Dirac string in Fig.1, it is the same as the contribution of a small cap in Fig.3b discussed below. We estimate the contribution of the cap by applying Stokes' theorem to a small circle surrounding the end of the Dirac string  in Fig.3b; we obtain the same contribution by ``blowing it up'' to the full outer surface in Fig.3b since we do not encounter any singularity by noting the fact that $r=0$ is not a singularity \eqref{singularity free}.  The  outside of the outer sphere in Fig.3b corresponds to the inside of the volume $V$ in Fig.1. We already know the contribution of the sphere in Fig.3b given by \eqref{total flux} and \eqref{integrated flux}, and thus the contribution of the small cap in Fig.1 is given by the minus of the contribution of the sphere  
\begin{eqnarray}\label{cap}
-\frac{e_{M}}{2} (1-\cos\frac{1}{2}\theta)
\end{eqnarray}
where we use a free value of $\theta$ without setting it at $\theta=\pi$ for the moment using $\Theta(\theta,\eta)=\frac{1}{2}\theta$ for $\eta=1$.  

The sum of \eqref{cylinder} and \eqref{cap} gives
\begin{eqnarray}
\frac{e_{M}}{2}(\cos\Theta(\theta,\eta)-\cos\frac{1}{2}\theta)-\frac{e_{M}}{2} (1-\cos\frac{1}{2}\theta)
=-\frac{e_{M}}{2}(1-\cos\Theta(\theta,\eta))
\end{eqnarray}
which gives $-e_{M}$ when one sets $\theta=\pi$ for $\eta>1$ and cancels the contribution from the outer surface \eqref{outer surface} in Fig.1, in agreement with Gauss' theorem \eqref{Gauss}.
 This analysis also shows that the Dirac string appears only when the net outgoing flux is nonvanishing.
 
The use of Stokes' theorem in the above analysis does not give an explicit evaluation of the small cap, but it shows the consistency of our analysis. The important property of the monopolelike object
\eqref{potential1} is that we have automatically the nonvanishing flux flowing through the surface of the thin tube surrounding the Dirac string \eqref{cylinder} including a small cap as described above, unlike the conventional monopole in \eqref{naive monopole} for which the incoming flux is introduced by hand to conserve the magnetic flux.

More formally, Stokes' theorem directly applied to an infinitesimally small circle $C$ surrounding the Dirac string in Fig.1 in the domain $\eta>1$  gives
\begin{eqnarray}\label{Stokes1}
\oint_{C} {\cal A}_{\varphi}r\sin\theta d\varphi =\int_{S^{\prime}}(\nabla\times {\cal A})\cdot d\vec{S}=e_{M}
\end{eqnarray}
using \eqref{total flux}.
This flux is regarded, depending on the choice of $S^{\prime}$, either as the flux flowing out of the volume $V$ which is explicitly given by \eqref{total flux} and \eqref{integrated flux} or the flux flowing into the volume $V$ through the surface of the tube covering the Dirac string including a cap at the origin of the Dirac string; the latter contribution is equal to the former contribution if one recalls the fact that no singularity (including $r=0$ as in  \eqref{singularity free}) exists inside the volume $V$ in Fig.1. We thus understand the vanishing of the left-hand side of \eqref{Gauss} in the case $\eta>1$. 
      
\begin{figure}[htb]
 \begin{center}
   \includegraphics*[width = 14.cm]{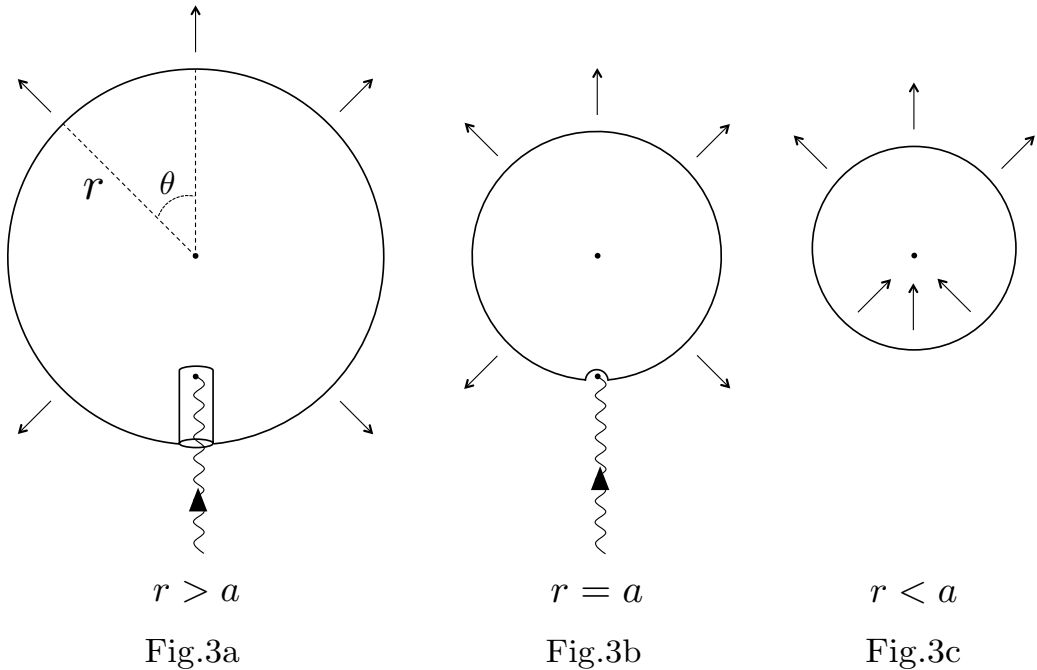}
   \caption{ 
   Pictures with varying radius $r$ with fixed $a$ (and thus varying $\eta=r/a$).  Geometrical center and the origin of the Dirac string denoted by a wavy line are displaced by the distance $r=a$.
The outgoing arrows indicate the flux in the radial direction on the sphere with fixed $r$.   
 The monopolelike potential vanishes at $r=0$ and thus the dot at the origin is for a visual aid and does not indicate any singularity. See also Figs.5 in \cite{Deguchi-Fujikawa1} in the case of a monopolelike object associated with Berry's phase.}\label{Fig_05a_b_c}
 \end{center}
\end{figure}

We next describe the pictures for varying $r$ with fixed $a$ in Figs.3a$\sim$ 3c using  the factor contained in the second expression in \eqref{total flux}
\begin{eqnarray}\label{differential flux}
\frac{e_{M}}{2}\Theta^{\prime}(\theta,\eta)\sin\Theta(\theta,\eta) 
\end{eqnarray}
and the movement of $\Theta(\theta,\eta)$ in Fig.2.
Both the Dirac monopolelike flux and the Dirac string indicated by a wavy line are seen when observed at $r>a$ in Fig.3a. In the transitional                                                                                                                                 domain at $r=a$ in Fig.3b, one recognizes both the out-going flux from an outer sphere and a small cap covering the origin of the Dirac string, although half of the strength of those in Fig.3a as is shown in \eqref{integrated flux}. No net outgoing flux and no Dirac string are observed when one comes closer to the monopole position $r<a$ in Fig.3c, which indicates that the flux is coming from a dipole.  The inward flux in Fig.3c is consistent with the negative signature of 
\begin{eqnarray}
\Theta^{\prime}(\theta,\eta)=\frac{\partial\Theta(\theta,\eta)}{\partial\theta}<0
\end{eqnarray}
in \eqref{differential flux} for $\eta<1$ and $\theta_{0}<\theta$, as is seen in Fig.2.  
From a point of view of the net outgoing flux, we thus have three distinguished configurations; the full flux with $e_{M}$ in Fig.3a and the half flux with $e_{M}/2$ in Fig.3b and then no net flux in Fig.3c, corresponding to $\Theta(\pi,\eta) = \pi$, $\pi/2$ and $0$, respectively, as already seen in \eqref{integrated flux}.

On the other hand, the surface $S$ on the left-hand side of Gauss' theorem \eqref{Gauss} does not cover any singularity (including $r=0$ which is not a singularity) and in this sense topologically trivial.  The Gauss' theorem \eqref{Gauss} is valid for a smooth decrease of $r$ starting with Fig.3a to Fig.3b and then to Fig.3c.  Our smoothness argument of topology change from a monopole to a dipole in the monopolelike potential is based on this smooth transition combined with the quantum mechanical interpretation of the Dirac string. 

We now follow the analyses of Dirac \cite{Dirac} and Wu and Yang\cite{Wu-Yang} to understand the crucial property of the Dirac string based on the notion of quantum mechanical states, which are described in the next section.

\section{Charge quantization and smooth topology change}
\subsection{Charge quantization}
We discuss the quantization of the electric charge in the presence of the present monopolelike potential using the notion of  quantum mechanical states.
If the Dirac string is not observable in the quantum mechanical sense when coupled to the electron, then we ignore it physically and identify a monopole.  This unobservability critically depends on the product of the electric charge of the electron and the magnetic charge of the monopolelike object and leads to the quantization of the electric charge~\cite{Dirac, Wu-Yang}, as will be discussed shortly.   Thus if the magnetic flux carried by the Dirac string satisfies the unobservability condition, we regard the monopolelike object as a physical monopole, and otherwise no physical monopole, namely, we regard a combination of the monopolelike object accompanied by the string as a physical entity.

We start with an analysis of the configuration with $\eta=r/a >1$ such as in Fig.3a.  The argument of Wu and Yang is to consider the singularity-free potentials in the upper and lower hemispheres        
\begin{eqnarray}
{\cal A}_{\varphi +} 
&=&  \frac{e_{M}}{4\pi r\sin\theta}(1 - \cos\Theta(\theta,\eta)),\nonumber\\
{\cal A}_{\varphi -} 
&=&  \frac{e_{M}}{4\pi r\sin\theta}(- 1 - \cos\Theta(\theta,\eta)),
\end{eqnarray}
using the potential in \eqref{potential1}. 
These two potentials are related by a gauge transformation 
\begin{eqnarray}
{\cal A}_{\varphi -}={\cal A}_{\varphi +} - \frac{\partial\Lambda}{r\sin\theta\partial \varphi}
\end{eqnarray}
with 
\begin{eqnarray}\label{gauge transformation}
\Lambda=\frac{e_{M}}{2\pi} \varphi.
\end{eqnarray}
The physical condition to be satisfied, which is related to the gauge invariance of the Aharonov-Bohm phase \cite{Aharonov-Bohm} , is 
\begin{eqnarray}
\exp[-\frac{ie}{c\hbar}\oint {\cal A}_{\varphi -}r\sin\theta d\varphi]&=&\exp[-\frac{ie}{c\hbar}\oint{\cal A}_{\varphi +}r\sin\theta d\varphi +\frac{ie}{c\hbar}\oint \frac{\partial\Lambda}{r\sin\theta\partial \varphi}r\sin\theta d\varphi]\nonumber\\
&=&\exp[-\frac{ie}{c\hbar}\oint{\cal A}_{\varphi +}r\sin\theta d\varphi].
\end{eqnarray}
This is satisfied if the condition $\exp[i e e_{M}/c\hbar]=1$ is satisfied by the gauge transformation \eqref{gauge transformation}, namely,  
\begin{eqnarray}\label{Dirac quantization condition}
e e_{M}/c\hbar=n\times 2\pi
\end{eqnarray}
with an integer $n$. 
It is confirmed that the present argument of gauge transformation is equivalent to the evaluation of the phase change induced by the Dirac string~\cite{Wu-Yang}, and \eqref{Dirac quantization condition} is the conventional quantization condition of Dirac \cite{Dirac}. (This electric charge quantization does not appear in Berry's phase where the magnetic charge by itself is quantized to be $e_{M}=2\pi\hbar$ \cite{Deguchi-Fujikawa1}.) 

In contrast, for the transitional domain $\eta=r/a =1$
such as in Fig.3b 
we have two potentials from \eqref{half sphere}  
\begin{eqnarray}
{\cal A}_{\varphi +}&=&\frac{e_{M}}{4r\sin\theta}(1-\cos\frac{1}{2}\theta)\nonumber\\
{\cal A}_{\varphi -}&=&\frac{e_{M}}{4r\sin\theta}(-\cos\frac{1}{2}\theta)
\end{eqnarray}
which are well-defined in the upper and lower hemispheres, respectively, and are related by the gauge transformation 
\begin{eqnarray}
{\cal A}_{\varphi -}={\cal A}_{\varphi +} - \frac{\partial\Lambda}{r\sin\theta\partial \varphi}
\end{eqnarray}
with 
\begin{eqnarray}
\Lambda=\frac{e_{M}}{4\pi} \varphi.
\end{eqnarray}
The physical condition 
\begin{eqnarray}
\exp[-\frac{ie}{c\hbar}\oint {\cal A}_{\varphi -}r\sin\theta d\varphi]&=&\exp[-\frac{ie}{c\hbar}\oint{\cal A}_{\varphi +}r\sin\theta d\varphi +\frac{ie}{c\hbar}\oint \frac{\partial\Lambda}{r\sin\theta\partial \varphi}r\sin\theta d\varphi]\nonumber\\
&=&\exp[-\frac{ie}{c\hbar}\oint{\cal A}_{\varphi +}r\sin\theta d\varphi]
\end{eqnarray}
then requires that the gauge transformation gives 
\begin{eqnarray}\label{half monopole phase1}
\exp[iee_{M}/2c\hbar]=\exp[in\pi ]=1
\end{eqnarray}
using the condition \eqref{Dirac quantization condition}.
For an even integer $n$ we can satisfy this condition, and thus we have a ``half-monopole'' with the magnetic charge $e_{M}/2$ as a physical monopole in Fig.3b.

On the other hand, for an odd integer $n$ in \eqref{Dirac quantization condition}, we have
\begin{eqnarray}\label{half monopole phase2}
\exp[iee_{M}/2c\hbar]=\exp[in\pi ]=-1.
\end{eqnarray}
This shows that the Dirac string associated with the half-monopole at the transitional domain $\eta=1$ with the magnetic charge $e_{M}/2$ is physically observable  for odd $n$; this phase of $\exp[i\pi]$ is the same as the Aharonov-Bohm phase of an electron in the magnetic field generated by the superconducting current of the Cooper pair in the experiment by Tonomura~\cite{Tonomura}.  In our criterion following the analysis of Wu and Yang \cite{Wu-Yang}, the Dirac string thus becomes a physical
observable.
The half-monopole is thus physical  only as a combination of  the monopolelike object, which generates the outgoing flux, and a Dirac string, although the Dirac string in the present case is actually defined only at the point $r=a$. Topologically, it is thus the same as the dipole for $\eta<1$ in Fig.3c for odd $n$.\\ 

\subsection { Smooth topology change}

The smooth topology change from a monopole to a dipole is a novel mechanism 
of topology change found in \cite{Deguchi-Fujikawa1}, which is also incorporated in the present model. The idea of the Dirac string plays a central role in this mechanism. From the point of view of the Gauss' theorem \eqref{Gauss}, all the configurations of our monopolelike object  are topologically the dipole as is seen in Fig.1 by recalling that $r=0$ is not a singularity for $a\neq 0$. 
Namely, the dipole which consists of the outgoing flux from the surface of the sphere and the incoming flux from the Dirac string has a trivial topology from the point of view of Gauss' theorem. But if one eliminates the Dirac string from the physical state space following the analysis of Dirac, and Wu and Yang, as described above, the dipole is now identified as a monopole.
The presence of a singularity at the origin is not required but the presence of the Dirac string is essential for the existence of the magnetic monopole; mathematically the Dirac string defines the monopole but the principle of quantum mechanics excludes the Dirac string from the space of physical states, leaving only the outgoing magnetic flux behind.

  Phenomenologically, the monopole and the dipole are very different if one looks at only the outgoing flux.  Mathematically, the transition from a monopole to a dipole when one approaches the monopole position as from Fig.3a to Fig.3c is nothing but the transition from a dipole to a dipole and thus the transition is smooth. This is the mechanism of the smooth topology change from a monopole to a dipole.   To realize the idea of a smooth transition from a monopole to a dipole, it is  important to have a finite gap (for $a\neq 0$) between the geometrical center of the monopolelike configuration and the origin of the Dirac string and that no singularity appears at the geometrical center $r=0$. 

We have already explained that the ordinary Dirac monopole is realized by setting $a=0$  first for any fixed $r$ in Fig.1. 
The Dirac string, which satisfies the Dirac quantization condition, is then stretching from the origin $\vec{r}=0$  to infinity for $a=0$. Thus there is no gap between the geometrical center and the origin of the Dirac string and, consequently, no topology change from a monopole to a dipole takes place. 

\section{Discussion and conclusion}

We have encountered an interesting monopolelike object in the parameter space in the analysis of an exactly solvable model of Berry's phase \cite{Deguchi-Fujikawa1}.  We here presented a translation of the monopolelike object in the parameter space to the monopolelike object in the ordinary space with a scale parameter $a$ replacing the period parameter in Berry's phase. In the long distance or infrared domain ($r\gg a$), our monopolelike potential looks like the conventional Dirac monopole with a Dirac string if the Dirac quantization condition is satisfied. But in the short distance or ultraviolet domain ($r\ll a$), the monopole disappears together with a Dirac string and becomes a dipole.

Both the conventional Dirac monopole and the present monopolelike object are mathematically the dipole in the sense that the outgoing magnetic flux is compensated for by the incoming flux through a Dirac string.  As Dirac pointed out, the Dirac string is quantum mechanically unobservable if the suitable charge quantization condition is satisfied \cite{Dirac}.
The difference between the two monopoles is that the  Dirac string originates from the origin of the coordinates in the conventional Dirac monopole as in Fig.1 with $a=0$ and thus looks like a genuine monopole if the Dirac string is excluded from the space of physical states.  In contrast, the monopolelike object we discussed is more like a dipole since the Dirac string originates from a point away from the origin of the coordinates as in Fig.1 with $a\neq 0$, and the magnetic flux in the increasing $\theta$ direction exists as in \eqref{magnetic flux} which, however, does not contribute to the net outgoing flux. Nevertheless, if one eliminates the Dirac string from the space of physical states in our monopolelike object by imposing Dirac's quantization condition, we have an object similar to the conventional Dirac monopole  when seen from far away. Moreover, we can explain the quantization of the electric charge in our monopolelike object also, although no point-like monopole exists for $a\neq 0$.  Technically, one might use our monopolelike object as a means to define the conventional monopole in a smooth manner in the limit $a\rightarrow 0$.

It is customary to mention the monopolelike object appearing in adiabatic Berry's phase \cite{Berry}. We here discussed the other way around, namely, the possible implications of Berry's phase on the magnetic monopole itself with a novel mechanism of smooth topology change from a monopole to a dipole.
It is hoped that the present analysis will stimulate the further analyses of the interesting subject of the Dirac magnetic monopole.
\\ 

One of us (KF) is supported in part by JSPS KAKENHI (Grant No.18K03633).

\end{document}